\documentclass[a4paper]{jpconf}
\newdimen\hdsize
\newcommand{\Mpc}{\>{\rm Mpc}}
\def\ltsima{$\; \buildrel < \over \sim \;$}
\def\lta{\lower.7ex\hbox{\ltsima}}

\def\rmb{{\rm b}}
\def\rmc{{\rm c}}
\def\rmd{{\rm d}}

\def\rmg{{\rm g}}
\def\rmh{{\rm h}}

\def\rmm{{\rm m}}

\def\rmp{{\rm p}}

\def\rms{{\rm s}}

\usepackage{graphicx}
\usepackage{epstopdf} 
\begin{document}
\title{The Galaxy-Dark Matter Connection:\\ A Cosmological Perspective}

\author{Surhud More$^{1}$, Frank van den Bosch$^{2}$, Marcello Cacciato$^{3}$,\\
 Anupreeta More$^{1}$, Houjun Mo$^{4}$, Xiaohu Yang$^{5}$}

\address{$^{1}$Kavli Institute for Cosmological Physics, University of Chicago,
933 East 56th Street, Chicago, IL 60637, USA }
\address{$^{2}$Astronomy Department, Yale University , Box 208101, New
Haven, CT 06520-8101, USA}
\address{$^{3}$Racah Institute of Physics, The Hebrew University, Jerusalem
91904, Israel}
\address{$^{4}$Department of Astronomy, University of Massachusetts, Amherst
MA 01003-9305}
\address{$^{5}$Key Laboratory for Research in Galaxies and Cosmology,
Shanghai Astronomical Observatory, the Partner Group of MPA, Nandan
Road 80, Shanghai 200030, China}

\ead{surhud@kicp.uchicago.edu}

\begin{abstract}
We present a method that uses observations of galaxies to
simultaneously constrain cosmological parameters and the galaxy-dark
matter connection (aka halo occupation statistics). The latter
describes how galaxies are distributed over dark matter haloes, and is
an imprint of the poorly understood physics of galaxy formation. A
generic problem of using galaxies to constrain cosmology is that
galaxies are a biased tracer of the mass distribution, and this bias
is generally unknown. The great advantage of simultaneously
constraining cosmology and halo occupation statistics is that this
effectively allows cosmological constraints marginalized over the uncertainties regarding galaxy bias.
Not only that, it also yields constraints on the galaxy-dark matter
connection, this time properly marginalized over cosmology, which is
of great value to inform theoretical models of galaxy formation. We
use a combination of the analytical halo model and the conditional
luminosity function to describe the galaxy-dark matter connection,
which we use to model the abundance, clustering and galaxy-galaxy
lensing properties of the galaxy population. We use a Fisher matrix
analysis to gauge the complementarity of these different observables,
and present some preliminary results from an analysis based on data
from the Sloan Digital Sky Survey. Our results are complementary to
and perfectly consistent with the results from the 7 year data release
of the WMAP mission, strengthening the case for a true `concordance'
cosmology.
\end{abstract}

\section{Introduction}

Cosmology has reached an important cross-road in the last couple of
decades, transitioning from a data-craved to a data-driven field of
science. The concordance cosmological picture of a Universe dominated
in energy density by dark energy and dark matter has emerged from a
vast number of cosmological investigations (see Figure $2$ in the
contribution by P.~J.~E.~Peebles in the current volume).  Ordinary
matter forms a fairly small component ($\sim 4.5\%$) of the energy
density of the Universe, with most of it present in the form of
intergalactic gas. The energy density of stars in galaxies has an
extremely negligible contribution to the energy budget. However,
unlike dark matter or dark energy, we can observe the star-light from
galaxies directly, and use galaxies as tracers of the underlying
matter density field to investigate the properties of the Universe.
Unfortunately, this connection between galaxies and (dark) matter is
complicated by the fact that galaxies are biased tracers of the mass
distribution. Although this `galaxy bias' is generally considered a
nuisance when trying to use galaxies to constrain cosmology, it also
contains a wealth of information regarding galaxy formation. After
all, it is the physics of galaxy formation that determines where, how
and with what efficiency galaxies form within the dark matter density
field. Therefore, ideally one would like to {\it simultaneously} solve
for cosmology and galaxy bias. In this paper, we present a method that
can do this, and show some preliminary results.

The overdensity of galaxies, $\delta_\rmg$, at a given position
$\vec{x}$, is related to the overdensity of matter, $\delta_\rmm$, at
that position, by a multiplicative term called the galaxy bias,
\begin{equation}
\delta_\rmg(\vec{x}) = b_\rmg\,\delta_\rmm(\vec{x})\,,
\end{equation}
and this implies that the power spectrum of the galaxy overdensity
field on a particular scale is related to the matter overdensity power
spectrum by
\begin{equation}
P_{\rm gg}(k) = b_\rmg^2(k)\,P_{\rm mm}(k)\,.
\end{equation}
In general, the galaxy bias defined in the above manner is expected to
be scale dependent \cite{Cacciato2012}. However, on large scales,
gravitation is the only relevant physics and galaxy bias is expected
to be scale free and equal to a constant. The shape of the galaxy
power spectrum on large scales, therefore, mimics the shape of the
matter power spectrum. The investigations of cosmological parameters,
in particular, the shape parameter $\Gamma=\Omega_m h$, have primarily
focussed on large scale precisely for this reason \cite{Tegmark2004,
Percival2007, Reid2010}.  However, it is also clear from the above
equations that on large scales, the amplitude of the power spectrum,
quantified by $\sigma_8$, is perfectly degenerate with the galaxy
bias, and large scale observations can only constrain the product
$b\sigma_8$ very well \cite{Tegmark2004}. The problem is further
complicated by the well known result that brighter galaxies cluster
more strongly than fainter galaxies, thus implying that the galaxy
bias is also luminosity dependent \cite{Norberg2001}.

It is crucial to understand why and how galaxies are biased with
respect to the matter distribution in order to break the degeneracy
between galaxy bias and the amplitude of the matter power spectrum.
The matter distribution in the Universe collapses to form bound clumps
of matter called halos. Since halos form preferentially at the peaks
of matter density field, halos themselves are biased tracers of the
underlying matter density field \cite{Mo1996}.  Galaxies form and
reside within these halos of dark matter, and therefore they inherit
the bias of their parent halos. Observations of the abundance of
galaxies \cite{Vale2006,Conroy2006}, the clustering of galaxies on
small scales \cite{Zehavi2005, Zehavi2011}, the gravitational lensing
signal due to the dark matter around galaxies
\cite{Mandelbaum2006,Leauthaud2012},
and the kinematics of satellite galaxies around halos
\cite{More2009a,More2009b,More2011a} can all provide important clues
regarding this ``galaxy-dark matter connection'' (i.e., what galaxies
resides in what halo). Using this information one can predict the
galaxy bias, both as function of scale and as function of galaxy
properties (e.g., luminosity). This allows one to break (some of) the
degeneracies between galaxy bias and cosmology, and thus to use the
observed distribution of galaxies to constrain cosmology.

Unfortunately, the clustering of galaxies is not sufficient to fully
break degeneracies. This is easy to understand. A generic prediction
of hierarchical formation scenarios is that more massive haloes are
more strongly clustered. Hence, the observed clustering strength of a
particular subset of galaxies (i.e., galaxies in a narrow luminosity
bin) on large scales, is a direct measure for the characteristic mass of their dark
matter haloes. However, different cosmologies predict different
clustering properties of the dark matter haloes. Consequently, the
galaxy-dark matter connection inferred from measurements of galaxy
clustering are strongly cosmology dependent \cite{vdB2003}. This
degeneracy can be broken using additional, independent constraints on
the galaxy-dark matter connection, such as those provided by satellite
kinematics or galaxy-galaxy lensing.

In this paper, we demonstrate the strength and complementarity of a
variety of galaxy observations to constrain cosmological parameters.
In particular, we show how observations of galaxy abundances, galaxy
clustering and galaxy-galaxy lensing, can be used to constrain
cosmological parameters such as the the matter density in units of the
critical density, $\Omega_\rmm$, and the amplitude of the power
spectrum of matter fluctuations, as characterized by $\sigma_8$.
We rely on the framework of the halo model to analytically predict
these observations. The halo model assumes that all the dark matter in
the Universe is partitioned over dark matter halos of different
sizes and masses \cite{Cooray2002}. The abundance and clustering of
these halos of dark matter is set by the underlying cosmological
parameters, and this dependence has been well calibrated with the use
of numerical simulations \cite{Tinker2008,Tinker2010}. A parametric
form of how galaxies populate halos, called the halo occupation
distribution function, can then be used to predict the abundance and
clustering of galaxies using the abundance and clustering of halos
\cite{Cacciato2009}. In this paper, we use a Fisher matrix analysis
to highlight the complementarity of using these different data sets,
and we present some preliminary results from an analysis based on
existing data.

\section{Data}
\label{sec:data}

We use data from the main galaxy sample of the Sloan Digital Sky
Survey (SDSS) \cite{York2000,Abazajian2009}. In particular, we use the
galaxy luminosity function, $\Phi(L)$, of \cite{Blanton2003}, the
projected two-point correlation functions, $w_\rmp(r_\rmp)$, for six
different luminosity bins and $0.2 h^{-1}\Mpc \lta r_\rmp \lta 40
h^{-1} \Mpc$, obtained by \cite{Zehavi2011}, and the excess surface
densities $\Delta \Sigma(r_\rmp)$, for the same six luminosity bins
but for $0.05 h^{-1}\Mpc \lta r_\rmp \lta 2 h^{-1} \Mpc$ from
\cite{Mandelbaum2006}.  The latter is proportional to the tangential
shear induced by the mass distribution associated with the galaxies in
question, and can be measured in the form of weak distortions of
background galaxies due to weak gravitational lensing (galaxy-galaxy
lensing). Our goal is to use a unified model that can describe all
these observables in terms of a simple parametric model, and to use
the existing data to simultaneously constrain cosmological parameters
and halo occupation statistics.

\section{Analytical framework}
\label{sec:frame}

We use the conditional luminosity function (CLF) to specify the halo
occupation distribution of galaxies \cite{Yang2003}. The CLF,
$\Phi(L|M)dL$, describes the average number of galaxies with
luminosity $L\pm dL/2$ that reside in a halo of mass $M$, and
consists of two components; one for central galaxies and the other
for satellites. Motivated by results obtained from a large SDSS
galaxy group catalogue \cite{Yang2008}, we assume that the CLF for
central galaxies is described by a log-normal distribution with a
logarithmic mean luminosity that depends on mass and a scatter
which we assume to be mass-independent. The dependence of the
logarithmic mean luminosity, $\tilde{L_\rmc}$ on halo mass is
parameterized using four central CLF parameters,
$L_0,\,M_1,\,\gamma_1$ and $\gamma_2$, and is given by
\begin{equation}
\tilde{L_\rmc}(M) = L_0 \frac{\left( M/M_1\right)^{\gamma_1}}{\left(
1+M/M_1\right)^{\gamma_1-\gamma_2}} \,.
\end{equation}
For the satellite component, we assume that it is well described
by a Schechter-like function
\begin{equation}
\Phi_\rms(L|M) = \Phi_* \left( \frac{L}{L_\rms}\right)^{\alpha_\rms} 
\exp\left[-\left(\frac{L}{L_\rms}\right)^2\right]\,,
\end{equation}
where the parameters, $L_\rms$, $\Phi_*$ and $\alpha_\rms$ are, in
general, functions of halo mass. Guided by the results of
\cite{Yang2008} based on a galaxy group catalog, we assume that
$L_\rms(M) = 0.562 \tilde{L}_\rmc(M)$ and that $\alpha_\rms$ is a
constant, independent of halo mass. The function $\log \Phi_\rms$ is
assumed to have a quadratic dependence on $\log M$, which is described
by three free parameters, $b_0$, $b_1$ and $b_2$.  Hence, the CLF,
which describes the halo occupation distribution as function of galaxy
luminosity, is described by a total of 9 free parameters. 

In addition to specifying the luminosity dependence of the halo
occupation distribution, we also need to specify the spatial
distribution of galaxies in dark matter halos. Throughout we assume
that central galaxies reside at the centers of their halos and that
the satellite galaxies follow the density distribution of dark matter
without any spatial bias. We have verified that this assumption has a
negliglible impact on our cosmological constraints.

Given the parameters of the central and satellite CLF, and the
cosmological parameters which set the abundance and clustering of
halos, we can predict all the observables that we wish to model. For
example, the luminosity function simply follows from multiplying
the average number of galaxies in a halo of mass $M$ by the number
density of halos of that mass, $n(M)$, and simply integrating this
product over all halo masses \cite{vdB2007},
\begin{equation}
\Phi(L)=\int \Phi(L|M) \, n(M) \, \rmd M\,.
\end{equation}
Similarly, the large scale bias of galaxies of luminosity $L$ can be
obtained by the following weighted average of the large scale bias of
dark matter halos, $b_\rmh(M)$, according to
\begin{equation}
b_\rmg(L) = \frac{ \int \Phi(L|M) \, b_\rmh(M) \, n(M) \, \rmd M }
{\int \Phi(L|M) \, n(M) \, \rmd M}\,,
\end{equation}

Because of the page-limits of these proceedings, we cannot provide the
detailed, analytical expresions that we use to calculate the observables 
$\Phi(L)$, $w_\rmp(r_\rmp)$, and $\Delta\Sigma(r_\rmp)$ for a given
model (i.e., cosmology plus CLF). These will be presented in van den
Bosch et al. (2012, in preparation). We emphasize, though, that our
implementation of the `halo model' \cite{Cooray2002} properly accounts
for (i) the scale dependence of halo bias, (ii) halo exclusion and
(iii) residual redshift space distortions that can affect the
determinations of galaxy bias \cite{More2011c}. Detailed tests using
realistic mock galaxy catalogs indicate that our analytical model is
accurate to better than 5 percent over the entire range of scales
covered by the data.

Throughout we adopt a `standard' flat $\Lambda$CDM cosmology (i.e.,
gravity is described by standard General Relativity, neutrino mass is
neglected, initial power spectrum is a single power-law, and dark
energy is modeled as Einstein's cosmological constant), which is
described by 5 cosmological parameters: the matter density parameter
$\Omega_\rmm$, the baryon density parameter $\Omega_\rmb$, the hubble
parameter $h$, the power law index $n_\rms$ and the parameter
$\sigma_8$.  Our goal is to constrain (subsets) of these cosmological
parameters, fully marginalizing over the galaxy-dark matter connection
as parameterized by our 9-parameter CLF model.
 
\section{Results}
\label{sec:cosmo}

\subsection{Fisher forecasts}

\begin{figure}[t]
\includegraphics[width=0.85\hdsize]{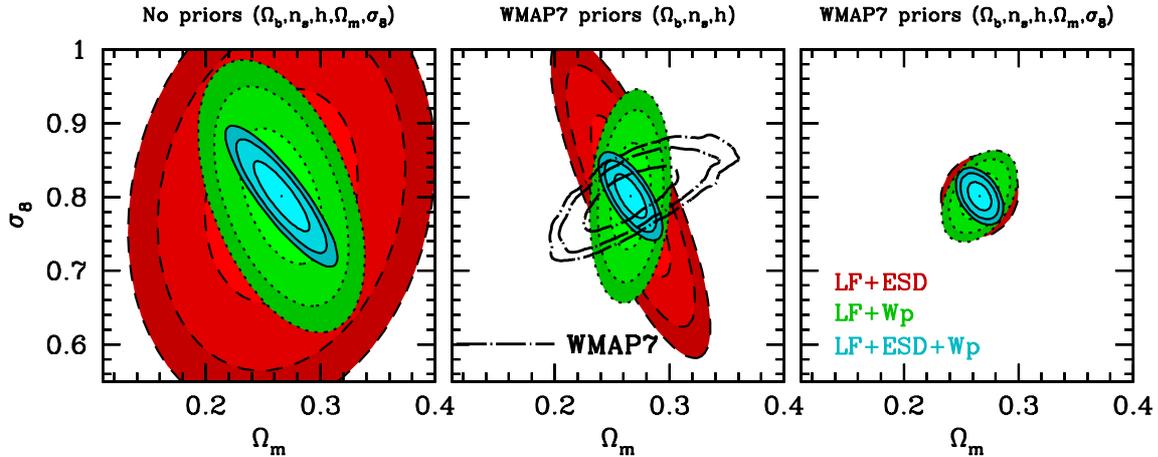}
\caption{\label{fisher}Fisher forecasts of 68, 95 and 99 percent
  confidence constraints on the cosmological parameters $\Omega_\rmm$
  and $\sigma_8$ when different combinations of the luminosity
  function (LF), the projected galaxy clustering (Wp) and the
  galaxy-galaxy lensing (ESD) data, as indicated in the legend, are
  analysed. Different panels show the effect of varying priors on
  cosmological parameters from analysis of the cosmic microwave
  background data. The dot-dashed contours in the middle panel show
  the constraints on $\Omega_\rmm$ and $\sigma_8$ from the 7 year data
  release of the WMAP mission, and are shown for comparison. For a
  definitive version, see More et al. (2012), in preparation.}
\end{figure}

In this section we use the Fisher information matrix in order to gauge
the accuracy with which constraints on the cosmological parameters
$\Omega_\rmm$ and $\sigma_8$ can be obtained given the current
accuracy of the observables that we wish to model.  Since we have
three different observables, the luminosity function, galaxy-galaxy
clustering and galaxy-galaxy lensing, we start by investigating how
each of these different data sets contribute to our constraining
power.

The different panels of Figure~\ref{fisher} show the 68, 95 and 99
percent confidence intervals that can be placed on the cosmological
parameters, $\Omega_\rmm$ and $\sigma_8$ under varying assumptions of
prior information from the 7 year analysis of the WMAP mission
\cite{Komatsu2011}. The left-hand panel assumes uninformative priors
on all of the cosmological parameters in our model. The dashed
contours are used to indicate the confidence levels when we perform a
joint analysis of the abundance of galaxies and the galaxy-galaxy
lensing signal around them. The constraints are fairly weak, in
particular, because the galaxy-galaxy lensing signal has only been
measured on fairly small scales ($r_\rmp \lta 2h^{-1} \Mpc$). This
results in a number of degeneracies between the cosmological
parameters and the CLF parameters such that $\Omega_\rmm$ and
$\sigma_8$ are only weakly constrained. The dotted contours show the
confidence contours obtained by combining the luminosity function with
the galaxy-galaxy clustering data. The constraints are significantly
tighter, and the improvement is largely due to the addition of
information on intermediate scales ($2 h^{-1}\Mpc \lta r_\rmp \lta 40
h^{-1}\Mpc$). Finally the solid contours show the result of a joint
analysis of all three observables. Even in the absence of prior
information, this joint analysis breaks a number of degeneracies that
are present in our model. The resulting cosmological constraints are
competitive with the existing constraints on these parameters,
demonstrating the potential power of this method.

The middle panel of Figure~\ref{fisher} shows the effect of adding
prior information on the secondary cosmological parameters,
$\Omega_b$, $n_s$ and $h$ from the 7 year WMAP results (hereafter
WMAP7). Notice how the addition of prior information can flip the
directions of degeneracies (compare the dashed contours in the
left-hand and the middle panels). The degeneracy between $\Omega_\rmm$
and $\sigma_8$ from our analysis is such that it runs perpendicular to
the WMAP7 constraints (shown by the dot-dashed contours). Adding these
WMAP7 constraints as additional priors on $\Omega_\rmm$ and
$\sigma_8$, further improves the constraints, as shown in the
right-hand panel.

The cosmological constraints presented above are also competitive
with those obtained from studies of the abundance of galaxy clusters
as a function of redshift. These galaxy clusters are detected either
via their X-ray emission \cite{Vikhlinin2009}, or as overdensities in
optical galaxy catalog \cite{Rozo2010}, or with the Sunyaev-Zel'dovich
effect \cite{Sehgal2011,Benson2011} and require extensive followup to
calibrate the cluster mass-observable relationship. Measurements of
cosmic shear have been used to obtain cosmological constraints, albeit
weaker, on $\Omega_m$ and $\sigma_8$ \cite{Hoekstra2006, Massey2007,
Lin2011, Huff2011}. Cosmological constraints have also been recently
obtained by analysing galaxy clustering and the mass-to-number ratio
on cluster scales by \cite{Tinker2012}. It is important to note that
all of these methods have very different systematics and are
complementary to each other and our method: they are all part of a
network of tests designed to validate the $\Lambda$CDM paradigm.

\subsection{Cosmological constraints}

\begin{figure}[t]
\includegraphics[width=15pc,angle=-90]{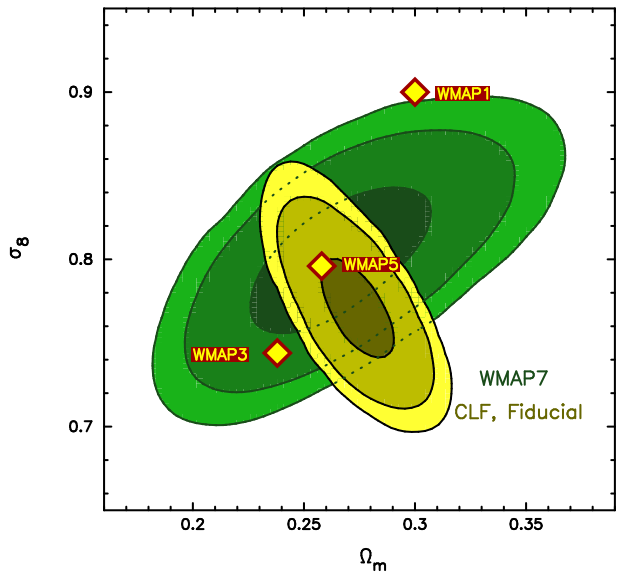}\hspace{5pc}%
\begin{minipage}[t]{15pc}\caption{\label{cosmo}68, 95 and 99 percent
    confidence limits on the cosmological parameters $\Omega_m$ and
    $\sigma_8$ from our analysis (shown in chrome yellow) compared
    with the confidence limits obtained by the analysis of the
      seven year data from the cosmic microwave background experiment
    WMAP (shown in green). For a more definitive version, see Cacciato
    et al.  (2012, in preparation).}
\end{minipage}
\vspace{2pc}
\end{figure}

We have carried out a joint analysis of all three observables, the
luminosity function, the projected galaxy clustering and the
galaxy-galaxy lensing signal, and obtained the posterior distribution
of our model parameters given these data. We use a Monte-Carlo Markov
chain to sample from posterior probability distribution of the
parameters. For our fiducial analysis, we impose priors on the
secondary cosmological parameters $\Omega_\rmb$, $n_\rms$ and $h$ and
completely uninformative priors on the parameters $\Omega_\rmm$ and
$\sigma_8$. Our model is able to fit the data sufficiently well with
$\chi^2$ per degree of freedom of the order of $2$. The fits to
the data will be presented in Cacciato et al. (2012, in preparation).
Preliminary results of our analysis, in the form of 68, 95 and 99
percent confidence contours, are shown in Fig.\ref{cosmo} and compared
to the WMAP7 results.

There are two points worth making. First of all, the constraints
obtained from our analysis are in remarkably good agreement with the
WMAP7 results, even though we have used no prior information on
$\Omega_\rmm$ and $\sigma_8$. The WMAP7 results are based on
observations of the microwave background at a very early time in the
Universe ($z\sim1080$) and primarily rest on the physics of
perturbations that can be treated with the help of linear perturbation
theory. The results from our analysis derive from galaxy observations
at redshift $z\sim0.1$ and are obtained by modelling extremely
non-linear scales, properly marginalizing over the uncertainties
  related to galaxy bias (i.e., the galaxy-dark matter connection).
The agreement in cosmological constraints obtained from these two
completely disjunct analyses is extremely striking and provides strong
support for the notion of a true `concordance' cosmology: clearly
  $\Lambda$CDM provides an excellent description of data over a large
range of scales and cosmic epochs.  Secondly, the constraints obtained
from our analysis are both competitive with and complementary to those
obtained by the WMAP analysis. This is also in agreement with the
complementarity expected from the Fisher forecast presented in the
previous subsection.

\section{Summary}
\label{sec:summary}

To summarize, observations of galaxies are an excellent way of probing
the underlying matter distribution in the Universe and thereby
obtaining precise constraints on the cosmological model. We have shown
that a joint analysis of the abundance of galaxies (characterized by
the galaxy luminosity function), the clustering of galaxies
(characterized by the projected two-point correlation functions), and
the clustering of dark matter around galaxies (characterized by
galaxy-galaxy lensing) can be a useful way to constrain the
cosmological parameters. We have modelled each of these observations
in the analytical framework of the halo model. The halo occupation
distribution of galaxies in our model was specified by the parametric
CLF model.

Using a Fisher matrix analysis, we have shown that the
cosmological information contained in the three observables described
above is complementary to each other and a joint analysis of these
datasets is able to break a number of degeneracies between the CLF
parameters and the cosmological parameters. We followed up our Fisher
forecast results, by constraining our model parameters using the
actual data. We have shown that the resulting constraints on the
cosmological parameters $\Omega_\rmm$ and $\sigma_8$ are in remarkable
agreement with constraints from the analysis of the WMAP data.  This
is yet another jewel in the crown of the $\Lambda$CDM model, which
continues to reign king.

We are currently exploring the use of our method to constrain
extensions of the $\Lambda$CDM model that include cosmological
parameters such as the neutrino density parameter, the dark energy
equation of state, non-gaussianity in the initial density fluctuations
and modifications to gravity. However, a large amount of work is still
required in order to calibrate the predictions of the abundance and
clustering of dark matter haloes in these alternative cosmologies.
Although many such calibrations for the extended $\Lambda$CDM already
exist in the literature, the halo mass definitions used in these
calibrations are often not suitable for use in the halo model
\cite{More2011b}. We expect to address these issues and extensions of
our methods in future work.

Finally, an interesting by-product of our analysis is a detailed,
statistical description of the galaxy-dark matter connection, as
parameterized by the CLF, fully marginalized over cosmological
uncertainties. This galaxy-dark matter connection is the outcome of
a large number of poorly understood astrophysical processes, such as
the formation of stars and the regulation of galaxy growth by
feedback, that shape the formation and evolution of galaxies in our
Universe. By constraining the CLF we are therefore constraining the
integral effect of these (and other) processes. Hence, the
constraints on the CLF parameters obtained from our analysis will be
of great value to inform theoretical models of galaxy formation
(both semi-analytic models and direct numerical simulations).

\section*{Acknowledgements}

This research has spanned more than three years during which the
affiliations of the different authors have changed. SM, FvdB and MC
would like to acknowledge support from the Max Planck Institute for
Astronomy and the University of Utah during the partial conduct of
this research. We also thank Jeremy Tinker, Alexie Leauthaud, Martin
White, Andrey Kravtsov, Eduardo Rozo, Matt Becker, Yin Li and Wayne Hu
for many interesting discussions and possible extensions of this
research work.

\end{document}